\documentclass[document]{emulateapj}

\shorttitle{Early afterglow of the short GRB090510}
\shortauthors{De Pasquale et al.}

\usepackage{amssymb}
\usepackage{amsmath}

\begin{document}

\title{{\it Swift} and {\it Fermi} observations of the early afterglow of
the short Gamma-Ray Burst 090510}

\author{M. De Pasquale\altaffilmark{1}\footnote{Corresponding authors: M. De
Pasquale: mdp@mssl.ucl.ac.uk; M. J. Page: mjp@mssl.ucl.ac.uk; K. Toma:
toma@astro.psu.edu; V. Pelassa: Veronique.Pelassa@lpta.in2p3.fr.}
, P. Schady\altaffilmark{1}, N.P.M. Kuin \altaffilmark{1}, M.J. Page \altaffilmark{1}, P.A. Curran \altaffilmark{1}, S. Zane\altaffilmark{1}, S.R. Oates\altaffilmark{1}, S.T. Holland\altaffilmark{2}, A.A. Breeveld\altaffilmark{1}, E. A. Hoversten\altaffilmark{3}, G. Chincarini\altaffilmark{4,5}, D. Grupe\altaffilmark{3}, {\it Fermi}/LAT and {\it Fermi}/GBM Collaborations}
\affil{$^1$Mullard Space Science Laboratory, University College London,
Holmbury St. Mary, Dorking, RH5 6NT, UK\\
$^2$ NASA/Goddard Space Flight Center, Greenbelt, MD 20771, USA\\
$^3$ Department of Astronomy \& Astrophysics, Pennsylvania State University, 525 Davey Lab, University Park, PA 16802, USA \\
$^4$ Universita~degli~Studi~di~Milano~Bicocca.~Piazza~della~Scienza,~3,~20126~Milano,~Italy\\
$^5$ Osservatorio~Astronomico~di~Brera~(INAF).~Via~E.~Bianchi~46,~23807~Merate~(LC),~Italy
}

\begin{abstract}

We present the observations of GRB090510 performed by the {\it Fermi}
Gamma-Ray Space Telescope and the {\it Swift} observatory. This is a bright,
short burst that shows an extended emission detected in the GeV
range. Furthermore, its optical emission initially rises, a feature so far observed only in long
bursts, while the X-ray flux shows an initial shallow decrease, followed
by a steeper decay. This exceptional behavior enables us to investigate
the physical properties of the GRB outflow, poorly known in short bursts. We
discuss internal shock and external shock models for the broadband
energy emission of this object.
\end{abstract}


\section{Introduction}

With the availability of a relatively large sample of Gamma-Ray Bursts (GRBs), we came to recognize that they comprise of two large classes \citep{kou93}: the so-called short-hard GRBs (duration $ \lesssim 2$~s) and the long-soft ones ($ \gtrsim 2$~s). There is now increasing consensus that the observed dichotomy  among long/short GRBs may indicate diverse initial physical conditions and progenitors. Long GRBs are associated with the demise of massive stars \citep{fer06}. Instead, short GRBs often occur in early-type galaxies \citep{zha09,geh09,fon09}. This supports their interpretation in terms of compact object mergers.

Crucial information on GRBs is revealed by their afterglows, which can be monitored by {\it Swift} \citep{geh04} in the optical and the X-ray range as soon as $\sim100$~s after the burst. In this paper, we present the study of the short GRB090510 with {\it Swift} and {\it Fermi} in a broad energy range, which extends from the optical up to a few GeV.

We report our observations and analysis in \S 2; in \S 3 we propose two different interpretations, in \S 4 we draw our conclusions. Hereafter, we use the conventions $X = 10^n X_n$ for cgs units and $F\propto t^{-\alpha} \nu^{-\beta}$, where $F$ is energy flux, $t$ is the time from the trigger of ${\it Swift}$ Burst Alert Telescope \citep{bar05a} and $\nu$ the frequency. Errors are reported at $1\sigma$, unless otherwise specified. We assume a cosmology in which $H_0=71$ km s$^{-1}$ Mpc$^{-1}$, $\Omega_m = 0.27$, $\Omega_\lambda=0.73$. All the fluxes, times and frequencies are measured in the observer's frame.

\section{Observations and data analysis.}

\subsection{BAT data}

At 00:23:00 UT, May 10th, 2009, BAT, which operates in the 15-350 keV range, triggered on GRB090510 \citep{hov09}. {\it Swift} slewed immediately to the burst. The duration was $T_{90} = 0.30 \pm 0.07$~s. Detailed analysis of the BAT data is shown in Ukwatta et al. (2009).

\subsection{XRT data}

The {\it Swift} X-ray Telescope \citep{bur05} began to observe the X-ray afterglow of GRB090510 at T+98~s. The lightcurve (Fig.~1) shows an initial slow flux decline. Observations were interrupted when the source entered the Earth constraint at T+1.9~ks. When they resumed, at T+5.1~ks, the flux was much lower. A broken powerlaw fit of the lightcurve gives as best fit parameters an early decay slope $\alpha_{X,1}=0.74\pm0.03$, break time $t_{X}=1.43^{+0.09} _{-0.15}$~ks, late decay slope $\alpha_{X,2}=2.18\pm
0.10$; $\chi^2=112$ with 77 degrees of fredom (d.o.f.), which is still marginally acceptable (chance probability $P=0.0054$).

\subsection{UVOT and other optical data}
The {\it Swift} Ultra Violet and Optical Telescope \citep{rom05,poo08} began settled exposures at T+97~s. The lightcurve of the optical afterglow, produced by renormalizing all individual filters to white, as described in Oates et al. (2009), is shown in Fig.~1. The optical emission rises until $\sim1.6$~ks, then decays. The optical lightcurve is well fitted ($\chi^2$ = 23.9 with 19 d.o.f.) by a broken powerlaw \citep{beu99} with a smooth break.
 The best fit parameters are: $\alpha_{Opt,1} = -0.50 ^{+0.11} _{-0.13}$; $t_{peak} = 1.58^{+0.46} _{-0.37}$~ks; $\alpha_{Opt,2} = 1.13^{+0.11} _{-0.10}$. Adding a constant does not improve the fit significantly, suggesting a small host galaxy contribution.
Very Large Telescope observations \citep{rau09} provide a spectroscopic redshift of $z=0.903$.
Using this redshift and the {\it Fermi} spectral parameters, the isotropic equivalent energy of GRB090510 is $E_{iso} = 1.08 \times 10^{53}$ ergs in the 10 keV - 30 GeV rest frame \citep{abd09}.

\subsection{{\it Fermi} data}

GRB090510 triggered both instruments onboard the {\it Fermi}
observatory \cite{gui09,ohp09}. The Gamma-ray Burst Monitor (GBM; 8 keV - 40 MeV) observed the burst
during the prompt emision phase, and after an autonomous repointing, the Large
Area Telescope (LAT; 20 MeV -  more than 300~GeV) began observations and detected
a long-lasting (up to 200s) high-energy (up to 4~GeV) emission. The analysis and interpretation of the
prompt emission is presented in Abdo et al.~(2009). Follow up observations lasted until
1500~s when the source was occulted by the Earth,
and resumed $\sim$ 3.5~ks later. The observation epochs are
defined in Table~\ref{tab:lat_ext_tab}. A time-resolved spectral 
analysis using diffuse events \cite{atw09} was performed
(Table~\ref{tab:lat_ext_tab}). The spectrum
shows no significant evolution, and it is well-fitted by a powerlaw with
energy index {\bf $\beta_\gamma =1.1 \pm 0.1$}. The lightcurve
shows no significant feature and is best fitted by a powerlaw.
The onset of the GBM emission (T+0.013~s) is a sensible reference
for this temporal fit and yields a decay index $\alpha_\gamma
=1.38\pm0.07$ ($\chi^2$/dof = 9.4/7) (Fig.~1). For the binned spectral analysis described in \S3.2,
transient events from an energy-dependent region of interest were used,
and front and back events were treated separately \citep{atw09}.

\section{Discussion} GRB090510 was a short burst with a relatively bright
afterglow, and $E_{iso}$ is among the highest for this class
\citep{gra09}. The early rise of the optical flux is so far unique in short
GRBs. More importantly, GRB090510 shows high energy emission up to
the GeV range, until T+200~s. 

Energetic short GRBs with optical transients, such as GRB050724
\citep{bar05b}, typically have extended emission (EE) detected by BAT
and XRT, following the hard emission spike \citep{tro08}. If GRB090510 had
occurred at $z=0.26$, as GRB050724, it would have produced a flux in the BAT
range of a few $10^{-9}$ ergs cm$^{-2}$ s$^{-1}$ until $\sim100$~s, 
 and we would have classified it as an EE-GRB.

The nature of this high energy emission is nevertheless not easy to
understand. EE often fades slowly for a few hundreds of seconds, then
vanishes with a slope which can be as fast as $\alpha \sim 7$; after this
sudden drop a late afterglow with a typical decay slope $\alpha
\sim 1.4$ is sometimes observed (e.g. GRB050724; GRB080123, Mangano et
al., 2008).
The fast decay and the extrapolation of the late afterglow back to
early epochs suggest that EE is not the onset of the late afterglow
\citep{nak07}. Furthermore, in a few cases where the EE is bright enough
to be studied in detail \citep{nob06}, it is found to have variations too
rapid to be explained with external shock models \citep{mer93}. EE might
instead indicate a declining activity of the GRB central engine
\citep{ros07,met08,per06,goa07}; once this activity ends, then falls
abruptly, and the forward shock (FS) emission prevails. In other cases,
however, the flux decay from the beginning of {\it Swift} observations
seems due to the usual FS mechanism, such as in GRB051221 \citep{bur06} and
GRB 061201 \citep{stra07}.

We propose and discuss separately 2 scenarios to explain the emission after the initial spike: in the first one, the
emission is due to both external FS and internal shock, \cite{rem94} while in the second the emission is due to FS alone.

\subsection{X-ray internal shock, optical external shock}

In the first scenario we assume that the initial X-ray (until the break at $\sim1.4$~ks) and $\gamma$-ray flux are IS emission, while the FS is responsible for the optical lightcurve and
the late X-ray flux. In particular, the optical rise may be due to the onset of FS emission, detected $10^2-10^3$~s after the trigger in long GRBs \citep{oat09}. This model can explain the different behavior of the early X-ray/LAT and optical lightcurves.

We constrain some physical properties of the FS blastwave. The initial Lorentz factor of the ejecta is $\Gamma_0 =
1.4 \times 10^2 E_{53} ^{1/8} n^{-1/8} t_{peak,3}^{-3/8}$ \citep{sar97},
where $E$ is the isotropic kinetic energy, $n$ the environment density
in $cm^{-3}$, and $t_{peak}$ is the peak time. The maximum FS flux is at the synchrotron characteristic
frequency $\nu_m$, and it is $F_{\nu_m} = 1.3 \times 10^4 E_{53}
\epsilon_{B,-2} ^{1/2} n^{1/2} \mu$Jy \citep{grs02}, where
$\epsilon_{B}$ is the fraction of internal energy in the magnetic field.

These parameter values must be consistent with $\Gamma_0 \gtrsim 1000$ \citep{abd09} and with the UVOT data, which give a 3$\sigma$ lower limit on $t_{peak} > 730$~s and a peak flux $F \simeq 100 \mu$Jy.  The first constraint can be written $ E_{53} n^{-1} > 2.6 \times 10^6 $. If  $\nu_m$ is just below the optical band, the constraint on the flux becomes $E_{53} \epsilon_{B,-2}^{1/2} n^{1/2} \simeq 7.7 \times 10^{-3}$. Assuming $\epsilon_{B,-2} \simeq 1$, the model is consistent with observations for $E_{53} \gtrsim 5.4 $, while $n \simeq 5.9 \times 10^{-5} E_{53}^{-2}$.

The XRT and LAT fluxes can be explained by IS synchrotron emission with $\nu_c< \nu_{Opt} < \nu_{sa} < \nu_X < \nu_m$, where $\nu_c$ is the synchrotron cooling frequency and $\nu_{sa}$ is the synchrotron self-absorption frequency \citep{gug03}. The synchrotron luminosity, estimated from the 100s spectral energy distribution (SED, Fig.~2), is  $L\simeq10^{50}$ erg s$^{-1}$. We find that for this value of $L$ and for $p=2.4$, $\epsilon_{e,-1} = 5.5$, $\epsilon_{B,-2} = 33$, $\Gamma=410$, $t_v = 3 \times 10^{-5}$~s, where $p$, $\epsilon_e$, $\Gamma$ and $t_v$ are the index of the powerlaw electron energy distribution, the fraction of energy given to electrons, the bulk Lorentz factor and the variability timescale respectively, we have $\nu_m \simeq 210$~keV, F(1.7~keV) $\simeq 75~\mu$Jy, which is within a factor $\sim3$ from the observations, and F(100 MeV) $\simeq 4.2 \times 10^{-3}~\mu$Jy, which is consistent within $2\sigma$ of the data. The cut-off energy for pair production is $h\nu_{\gamma\gamma} \simeq 1.6$~GeV, thus allowing the late emission of $\sim1$~GeV photons. IS does not produce detectable emission in the optical, since this is below $\nu_{sa} \simeq 0.32$~keV.

Compared with the scenario presented in \S3.2, this model has the advantage of not requiring extreme values of $\Gamma_0$ to explain an early (few seconds) FS emission onset in a low density environment expected for a short burst. However, it needs some fine tuning of parameters. The optical rise slope $\alpha_{Opt,1}=-0.5$ is shallower than that expected at the FS onset ($\alpha=-2$), although similar slow rises have been observed \citep{oat09}. It is possible that the onset of our observations caught the end of this steep rise phase, when the afterglow was turning over to a decay. Another possible problem is that the required density appears very low. We note that the observed slope would be expected if $\nu_m$ were crossing the optical band and, in general, broad FS onset rises are also expected for outflows observed off-axis \citep{pav08}. However, a bright and hard event such as GRB090510 is difficult to reconcile with the latter scenario, which predicts soft and dim prompt emission \citep{yam03}.

\subsection{Optical, X-ray and GeV emission from external shock.}

A second possibility is that the afterglow of GRB090510, including the
emission detected by LAT, is entirely produced by the FS propagating
in a constant density medium \cite{sar98}. According to the model,
the broad afterglow spectrum\footnote{The self-absorption frequency is
not relevant in this  study.} consists of three segments: a low-energy tail, of spectral slope
$\beta_1 = -1/3$; another segment, for $\nu_m < \nu < \nu_c$, where
$\beta_2
= (p-1)/2$; blueward of $\nu_c$, the third segment has $\beta_3 = p/2$.
For comparison with this spectral template, we built up 5 SEDs, at 100~s, 150~s, 1~ks, 7~ks, and
12~ks (Fig.~3), all including UVOT and XRT data, and LAT data were also included in the first SED.
LAT data were accumulated between 10~s and 200~s (i.e. well after the end of the prompt emission
seen in the GBM), and renormalized to 100~s using the decay index of $\alpha_{\gamma}=1.38$. We fitted the SEDs simultaneously with a double broken powerlaw model, forcing $\beta_1 = -1/3$ and $\beta_3=\beta_2 + 1/2$. We allowed the breaks to vary and
Galactic and host extinction were accounted for. The result is acceptable ($\chi^2$/d.o.f.=110.3/83) and is shown in
Fig.~3 and Table~\ref{tab:SEDs_FIT}. The FS alone could successfully describe the spectrum over 9 decades of frequency. A break between X-ray and $\gamma$-ray ranges is fitted, at $E^{b} _{2} \simeq 300$~MeV, but not constrained. It is studied more precisely by fitting the 100~s SED alone, freezing $N_H$ and $E(B-V)$
at the 5 SEDs fit results and leaving $\beta_2$ and
$\beta_3$ free to vary (see Table~\ref{tab:SEDs_FIT}). This fit fulfills
the relation $\beta_3\,=\,\beta_2\,+\,1/2$ (at 1.3 $\sigma$) and a
significant break is found (3.6 $\sigma$), although it yields a slightly
harder $\beta_2$ (1.8 $\sigma$) than the 5 SED fit shown in Fig.~3. The LAT emission shows no spectral
evolution, even at early times. Therefore, to better characterize the
high-energy spectrum, the SED at 100~s was rebuilt including LAT data
between 0.38~s and 200~s (see Table~\ref{tab:SEDs_FIT} and Fig.~2). A
significant break ($>4.5\sigma$ including systematics) between 10 and 133~MeV was found.
However, including this selection of LAT data in the
5 SED fit yields a worse fit ($\chi^2$/dof = 125.3/83)
than that shown in Fig.~3.

In this FS interpretation, the initial increase of the optical emission is due to $\nu_m$ approaching the optical band. The X-ray is already decaying because it lies above $\nu_m$. In order to verify whether the required physical parameters are
plausible, we impose the following constraints: i) $F(1\;{\rm ks})
\simeq 2.2 \;\mu$Jy at $10^{18}\;$Hz and ii) $\nu_m (1\;{\rm ks}) \simeq
10^{16}\;$Hz. Adopting the expressions for $F_{\nu_m}$, $\nu_m$, and $\nu_c$
from Granot \& Sari (2002), the constraints above lead to the following equations: $\epsilon_{B,-2} \simeq
14\;E_{53}^{-1/2} \epsilon_{e,-1}^{-4} \xi_p^{-4} \nu_{os}^2$, $n \simeq
1.5 \times 10^{-6}\;(1330)^{p-2.5} E_{53}^{-1} \epsilon_{e,-1}^4 \xi_p^4
\nu_{os}^{-p-1}$, where $\xi_p = 3(p-2)/(p-1)$ and $\nu_{os} = \nu_m
(1\;{\rm ks})/10^{16}\;{\rm Hz}$. We verified that the
synchrotron self-Compton cooling is not significant for $t \leq
1.5\;$ks and $\nu_c (1.5\;{\rm ks}) > 10^{18}\;$Hz for a
reasonable range of parameters \citep{nak09}. A very low, but not
implausible, density is suggested.

The flux at 100MeV is
\begin{equation}
\begin{matrix}
F_{\nu>\nu_c} \simeq 2.4\times10^{-3} (2.2\times10^{-3})^{(p-2.5)}
E_{53} \epsilon_{e,-1} \\ \hspace{6mm} (t/100s)^{(2-3p)/4} (h\nu/{\rm 100MeV})^{-p/2}
\xi_p \nu_{os}^{(p-2)/2} \mu {\rm Jy}
\end{matrix}
\end{equation}
This is consistent with the LAT data at 100s, provided
$E_{53} \epsilon_{e,-1} \simeq 5$, $p \approx 2.5$, $\xi_p \approx 1$,
and $\nu_{os} \approx 1$. For these parameters, $\nu_c \ll 4$GeV
at $t \gtrsim 1$~s, so that the flux in the LAT energy range is
approximated to be $F \propto t^{(2-3p)/4} \sim t^{-1.4}$ at
$t \gtrsim 1$~s, consistent with the LAT light curve.
We note that $\Gamma_0 > 5800 E_{53}^{1/8} n_{-4}^{-1/8}$ is required
for the FS onset time to be $\lesssim 1$~s.

In summary, the spectral properties of GRB 090510 could be explained by a
simple FS model.
We note, though, that
this simple model is hard to reconcile with some of the observed temporal
properties. Firstly, it predicts an X-ray decay index before the break of
$\alpha = 3\beta_2/2 = 1.16 \pm 0.06$, clearly inconsistent with the
observed $\alpha_{X,1} = 0.74 \pm 0.03$. Secondly, if the X-ray break at
$t = 1.4\;$ks is attributed to a jet break \citep{sar99} and, after
this time, optical and X-ray lie on the same spectral segment, then the
asymptotic optical decay index should be the same as $\alpha_{X,2} = 2.18
\pm 0.10$. However, a fit of the whole lightcurve with a smooth broken
power-law (\S 2.3), adding a constant (as a host galaxy contribution), gives an
asymptotic decay slope $\alpha_{Opt,2} = 1.13^{+0.17}_{-0.09}$,
incompatible with the X-ray decay. Finally, although the error bars are quite large, we notice that, taken at face value, the slope of the UVOT spectrum in the $1\;$ks SED is negative (Fig.~3), suggesting that
$\nu_m$ may be already below the optical at that epoch.

The above mentioned flaws imply that the simple FS model is
not viable to explain the properties of GRB090510. However, this model
relies on highly idealized assumptions, and it is known that several GRB
afterglows do not strictly follow its simple predictions.
Plausible effects that may affect the predictions and ease the comparison with GBR090510 are:

- a phase of energy injection \citep{sam00}, or an evolution of the
microphysical parameters of the blast wave \citep{pan06}; both may cause
an early shallow decay of the X-ray flux;

- the transit of $\nu_m$ slightly after the jet break, which could explain
a shallow late optical decay.

With the present data, we are unable to distinguish between energy injection and microphysical parameter evolution. As for the X-ray decay post jet break, hydrodynamical simulations show that initially the jet decay slope can reach $\alpha \simeq 3$ \citep{gra06}. Both the processes of energy injection and parameter evolution are capable of keeping the decay shallower, so that a late X-ray decay slope of $\alpha_{X,2} = 2.18 \pm 0.10$ could be achieved. Therefore, with some extensions, the FS model could arrange the temporal properties, although some fine tunings would be needed.

\section{Conclusions}

We have reported the {\it Swift} and {\it Fermi} observations of the short
GRB090510, an event endowed with bright prompt and afterglow emission, and
detected in the GeV range up to 200~s after the trigger. The initial X-ray
emission shows a slow decay up to $\simeq 1.4$~ks, after which it quickly
drops. The optical flux peaks at $\simeq 1.6$~ks. We have explored two scenarios to explain the observed behaviors.\\
  In the first scenario, the early flux detected by XRT and LAT is due to IS,
while the optical rise is the onset of FS emission or the transit of $\nu_m$. This
interpretation does not require extremely high values of Lorentz factor,
should the density of the environment be very low. We also find that
reasonable values for the physical parameters can lead to the observed
properties, which might favour the model, although some fine tuning is
necessary. The second scenario assumes that the FS produces the full spectrum of
the emission, observed from the optical to the GeV band. The $\gamma$-ray, X-ray and optical spectrum can be reproduced by the template FS spectral
model and the required physical parameters are plausible.
 Although the simple FS model fails to reproduce the observed temporal
behavior, extensions of this model could accommodate the temporal
mismatch. In order to identify the origin of the GeV component of GRBs
like 090510, more case studies will be necessary. Fortunately, we have very promising
prospects for other simultaneous {\it Fermi} and {\it
Swift} observations of short GRBs, which will provide us with more measurements
to shed light on the properties of this class of events.

\acknowledgements
The {\it Fermi} LAT Collaboration acknowledges support from a number of agencies and institutes for both development and the operation of the LAT as well as scientific data analysis. These include NASA and DOE in the United States, CEA/Irfu and IN2P3/CNRS in France, ASI and INFN in Italy, MEXT, KEK, and JAXA in Japan, and the K.~A.~Wallenberg Foundation, the Swedish Research Council and the National Space Board in Sweden. Additional support from INAF in Italy and CNES in France for science analysis during the operations phase is also gratefully acknowledged.
SZ aknowledges STFC support. This work used data supplied by the UK Swift SDC at the University of Leicester.

\clearpage

\begin{table}
\begin{center}
\begin{tabular}{|c|c|c|c|}
\hline
Time & Test & Energy & Photon flux above 100$\,$MeV \\
bins (s) & Statistic & index & ($ph.cm^{-2}.s^{-1}$) \\
\hline
\hline
1: 0.38 -- 0.48 & 208.2 & 0.85$^{- 0.26}_{+ 0.30}$ & 2.49$^{+
  1.13}_{- 0.84}$ $10^{-2}$ \\
2: 0.48 -- 0.92 & 541.8 & 1.20$^{- 0.20}_{+ 0.22}$ & 1.89$^{+
  0.46}_{- 0.39}$ $10^{-2}$ \\
3: 0.92 -- 1.5 & 192.1 & 0.93$^{- 0.26}_{+ 0.30}$ & 5.7$^{+ 2.4}_{-
  1.8}$ $10^{-3}$ \\
4: 1.5 -- 2.5 & 301.7 & 1.41$^{- 0.28}_{+ 0.31}$ & 6.4$^{+ 2.0}_{-
  1.6}$ $10^{-3}$ \\
5: 2.5 -- 5.5 & 163. & 0.76$^{- 0.22}_{+ 0.26}$ & 8.4$^{+ 3.6}_{-
  2.7}$ $10^{-4}$ \\
6: 5.5 -- 11.5 & 58. & 0.86$^{- 0.35}_{+ 0.44}$ & 2.0$^{+ 1.4}_{-
  0.9}$ $10^{-4}$ \\
7: 11.5 -- 37. & 71. & 2.27$^{- 0.59}_{+ 0.70}$ & 1.67$^{+ 0.82}_{-
  0.59}$ $10^{-4}$ \\
8: 37. -- 69.5 & 59.9 & 0.85$^{- 0.32}_{+ 0.39}$ & 4.4$^{+ 2.6}_{-
  1.8}$ $10^{-5}$ \\
9: 69.5 -- 200. & 43. & 1.74$^{- 0.58}_{+ 0.71}$ & 1.6$^{+ 1.0}_{-
  0.7}$ $10^{-5}$ \\
\hline
\hline
10: 200. -- 400. & $\sim0.$ & 1.1 / 0.5 / 2.5 & $<$ 4.7 / 3.2 / 7.9 $10^{-6}$\\
11: 400. -- 800. & $\sim0.$ & 1.1 / 0.5 / 2.5 & $<$ 2.3 / 1.6 / 3.9 $10^{-6}$\\
12: 800. -- 1500. & $\sim0.$ & 1.1 / 0.5 / 2.5 & $<$ 2.1 / 1.2 / 3.8 $10^{-6}$\\
13: 4200. -- 7200. & $\sim0.$ & 1.1 / 0.5 / 2.5 & $<$ 0.48 / 0.3 / 1.0 $10^{-6}$\\
14: 10150. -- 13000. & $\sim0.$ & 1.1 / 0.5 / 2.5 & $<$ 0.47 / 0.3 / 0.9 $10^{-6}$\\
15: 15800. -- 18500. & $\sim0.$ & 1.1 / 0.5 / 2.5 & $<$ 0.52 / 0.3 / 1.0 $10^{-6}$\\
\hline
\hline
\end{tabular}
\end{center}
\caption{\label{tab:lat_ext_tab}
LAT time-resolved spectroscopy. The source test 
statistic was defined as twice the difference of the unbinned
log-likelihood between the null hypothesis (background only) and the
alternative hypothesis (presence of a source). Bottom table shows 95\% confidence level upper
  limits on flux for different energy indeces.}
\end{table}

\begin{table}
\begin{center}
\begin{tabular}{|c|c|c|c|}
\hline
SED & 5 SED & 100~s & 100~s \\
LAT dataset (s) & 10. -- 200. & 10. -- 200. & 0.38 -- 200. \\
\hline
$N_H$ ($\times$ 10$^{21}$ cm$^2$) & 1.52 $^{+0.33} _{-0.30}$ & 1.52 (fixed) & 1.52 (fixed) \\
$E(B-V)$ (mag) & 0.000 $^{+0.005}_{-0.000}$ & 0. (fixed) & 0. (fixed) \\
$E_1^b$ (keV) & $\left\{ \begin{array}{ll}
    0.43 ^{+0.10}_{-0.07} &\mbox{(100~s)} \\
    0.17 ^{+0.03}_{-0.02} &\mbox{(150~s)} \\
    0.037 ^{+0.005}_{-0.007} &\mbox{(1000~s)} \\
    < 0.001 &\mbox{(7000~s)} \\
    < 0.01 &\mbox{(12000~s)} \\
  \end{array}\right.$ & 0.31 $^{+0.05}_{-0.06}$ & 0.31 $^{+0.06}_{-0.05}$ \\
$\beta_2$ & 0.77 $\pm$ 0.04 & 0.61 $^{+0.06}_{-0.10}$ & 0.62 $^{+0.08}_{-0.06}$ \\
$E_2^b$ (MeV) & $\simeq$ 300 (100~s) & $[20-135]$ & $[10-133]$ \\
$\beta_3$ & $\beta_2\,+\,1/2$ & 1.44 $^{+0.26}_{-0.22}$ & 1.14 $^{+0.10}_{-0.09}$ \\
\hline
\end{tabular}
\end{center}
\caption{\label{tab:SEDs_FIT}
Best fit parameters obtained by fitting the 5 SEDs simultaneously or
the SED at 100~s only with a double broken powerlaw model (see text). $N_H$ and $E(B-V)$ are host absorption and extinction respectively; $E_1^b$ and $E_2^b$ are the two break
energies calculated at the epoch in parenthesis.}
\end{table}

\clearpage

\begin{figure}
\begin{center}
\includegraphics[width=0.65\linewidth] {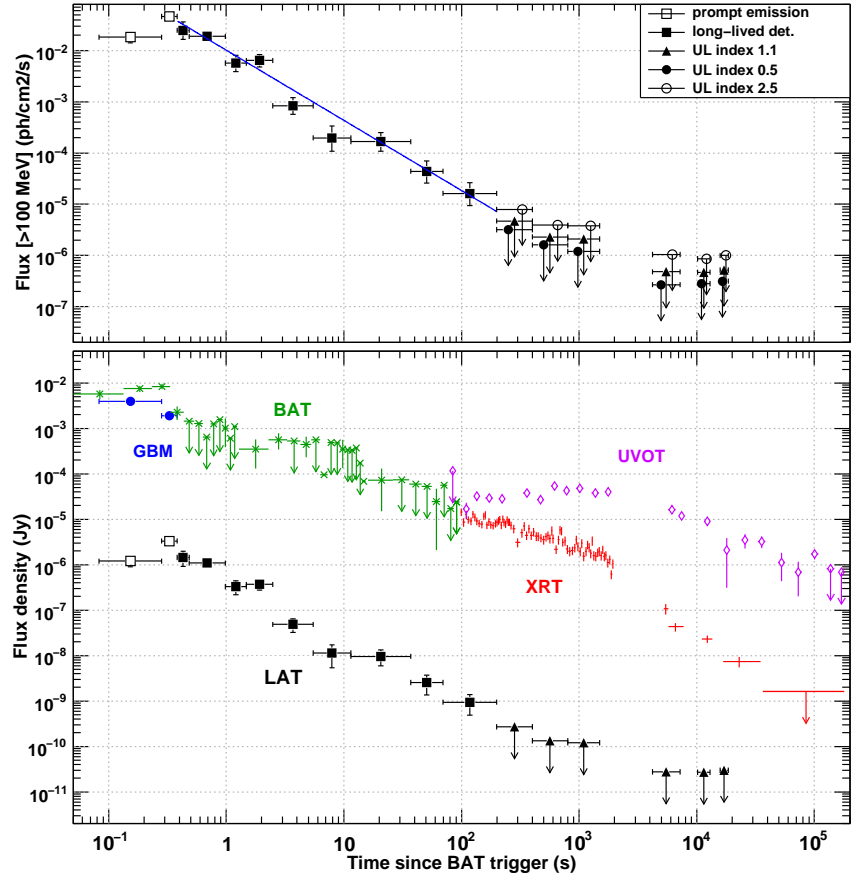}
\end{center}

\caption{ {\bf Top: } LAT flux above 100 MeV and best fit to the
flux decay (line).
{\bf Bottom: }energy flux densities averaged in the observed
energy bands: BAT (15 keV -- 350 keV, stars); XRT
(0.2 keV -- 10 keV, crosses); UVOT renormalised to white
(diamonds);
LAT (100 MeV -- 4 GeV, filled squares; the average spectral index
was used to convert from photon to energy flux) with
upper limits for $\beta\,=\,1.1$ (triangles). The prompt emission is shown for
comparison: GBM (8 keV -- 1 MeV, circles), LAT (100 MeV -- 4 GeV, empty squares). XRT lightcurve is obtained as in Evans et al. (2007, 2009).\\
All data are shown with 68\% error bars or 95\% confidence level upper limits.}
\label{fig:MW_LC}
\end{figure}

\clearpage

\begin{figure}
\begin{center}
\includegraphics[width=0.7\linewidth,angle=270]{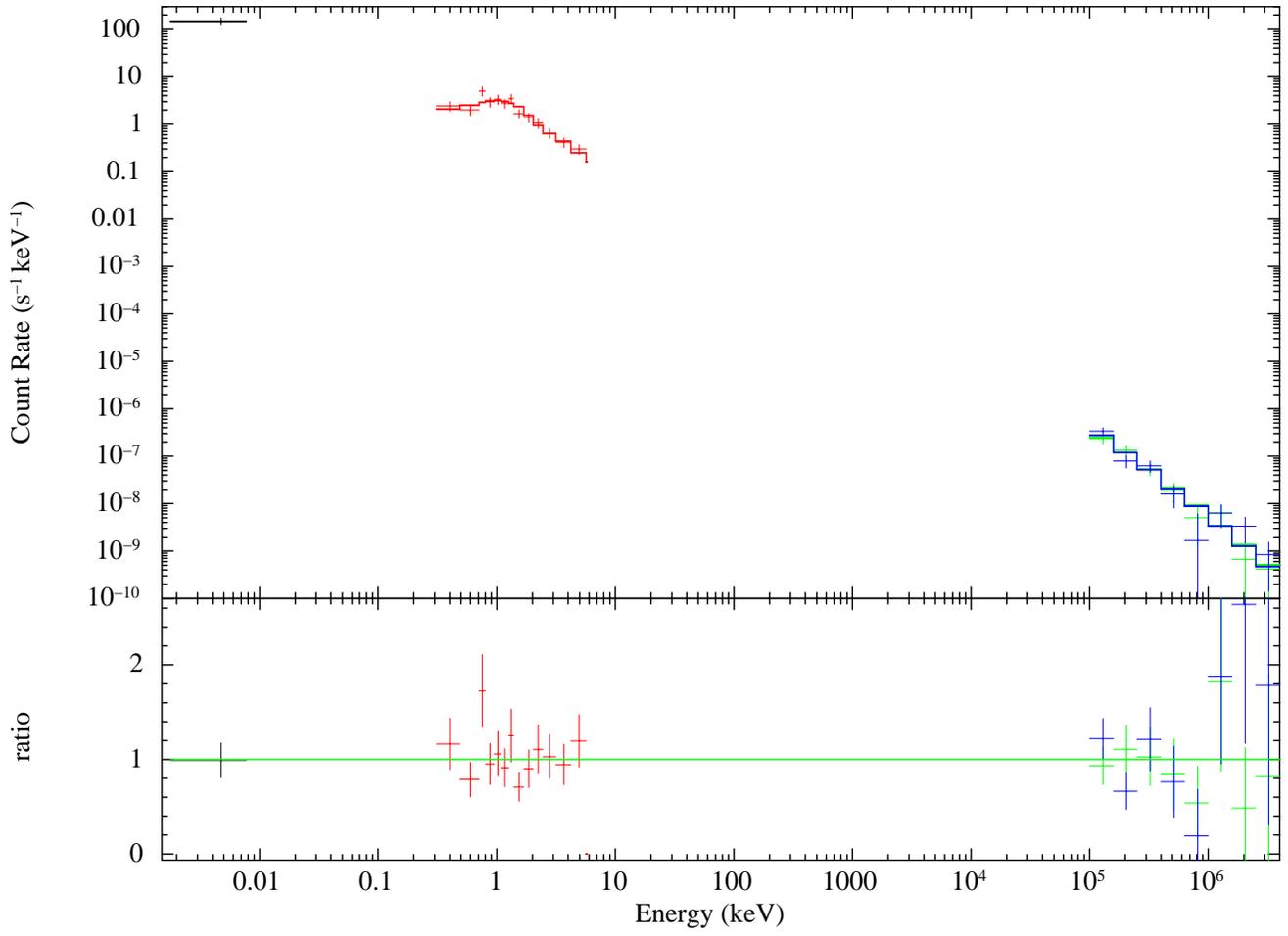}
\end{center}
\caption{\label{fig:xrt_lat_sed}
  UVOT-XRT-LAT count spectrum, with best fit and residuals shown (see text)
  }
\end{figure}

\clearpage

\begin{figure}
\begin{center}
\includegraphics[angle=0, width=0.7\linewidth]{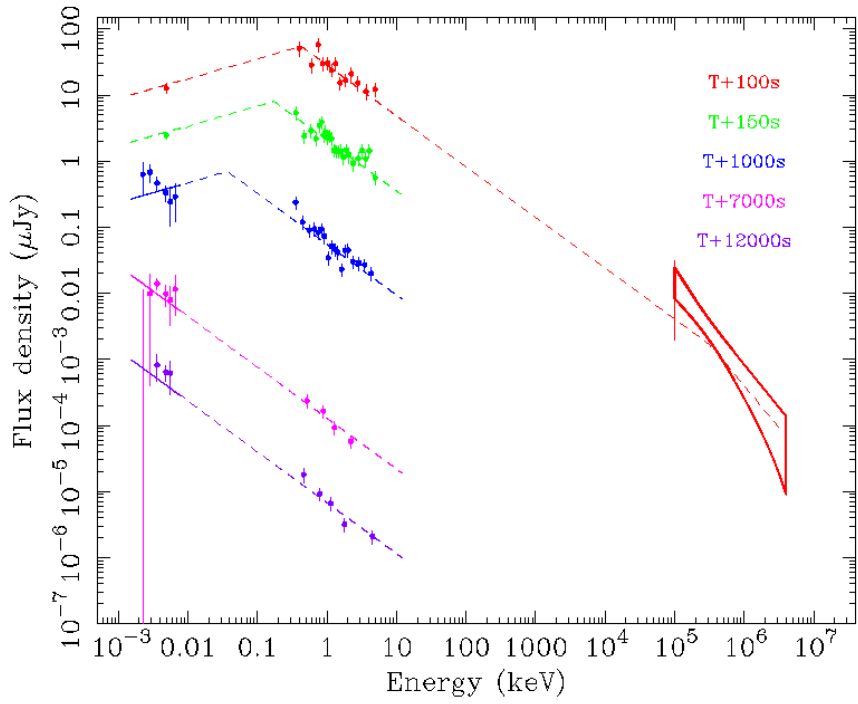}
\end{center}
\caption{\label{fig:SEDs}
 UVOT-XRT-LAT SEDs at different epochs, with the best fit shown (see text). The butterfly at 100s indicates the 68\% confidence level region for the LAT flux, obtained from an unbinned likelihood analysis (95\% error bar at 100~MeV is shown). Successive SEDs in time order are rescaled by 1:1, 1:10, 1:100, 1:1000, 1:10000.}
\end{figure}


\begin{thebibliography}{}
\bibitem[{Abdo et al. }{2009}]{abd09} Abdo A.A., et al. 2009, in preparation
\bibitem[{Atwood et al. }{2009}]{atw09} Atwood, W., Abdo, A. A., Ackermann, M., et al.  2009 ApJ, 697, 1071
\bibitem[{BAT; Barthelmy et al. }{2005a}]{bar05a} Barthelmy, S.D.,  Barbier, L.M., Cummings, J.R., et al. 2005, SSRv, 120, 143
\bibitem[{Barthelmy et al. }{2005b}]{bar05b} Barthelmy, S.D., Chincarini, G., Burrows, D.N., et al. 2005, Nature, 438, 944
\bibitem[{Beuermann et al. }{1999}]{beu99} Beuermann, K., Hessmann, F.V., Reinsch K. et al. 1999, A\&A, 352, 26
\bibitem[{XRT; 0.3-10~keV; Burrows et al. }{2005}]{bur05} Burrows, D.N., Hill, J.E., Nousek, J.A., et al. 2005, SSRv, 120, 165
\bibitem[{Burrows et al. }{2006}]{bur06} Burrows, D.N., Grupe, D., Capalbi, M., et al. 2006, \apj, 653, 468
\bibitem[{Evans et al. }{2007}]{eva07} Evans, P.A., Beardmore, A.P., Page, K.L. et al. 2007, A\&A 469, 379-385
\bibitem[{Evans et al. }{2009}]{eva09} Evans P.A., Beardmore, A.P., Page K.L., et al. 2009, \mnras~submitted, arXiv:08123662
\bibitem[{Ferrero et al. }{2006}]{fer06} Ferrero P., Kann, D.A., Zeh, A., et al. 2006, A\&A 457, 857
\bibitem[{Fong et al. }{2009}]{fon09} Fong, W., Berger, E., \& Fox, D. B. 2009, \apj, submitted, arXiv:0909.1804
\bibitem[{Gehrels et al. }{2004}]{geh04} Gehrels, N., Chincarini, G., Giommi P., et al. 2004, \apj, 611, 1055
\bibitem[{Gehrels et al. }{2009}]{geh09} Gehrels, N., Ramirez-Ruiz, E., \& Fox, D. S. 2009, ARAA, 47, 567
\bibitem[{Graham et al. }{2009}]{gra09} Graham, J. F., Fruchter, A. S., Levan, A. J., et al. 2009, \apj, 698, 1620
\bibitem[{Goad et al. }{2007}]{goa07} Goad, M.R., Page K.L., Godet, O., et al 2007, A\&A, 468, 103
\bibitem[{Granot }{2007}]{gra06} Granot J. 2007, RMAC, 27, 140
\bibitem[{Granot \& Sari }{2002}]{grs02} Granot, J., \& Sari., R. 2002, \apj, 568, 820
\bibitem[{Guetta \& Granot }{2003}]{gug03} Guetta, D., \& Granot, J. 2003, \apj, 585, 885
\bibitem[{Guiriec et al. }{2009}]{gui09} Guiriec, S., Connaughton, V. \& Briggs M. 2009, GCN 9336
\bibitem[{Hoversten et al. }{2009}]{hov09} Hoversten,  E.A., Krimm, H.A., Grupe, D. et al. 2009, GCN 9331
\bibitem[{Kouveliotou et al. } {1993}]{kou93} Kouveliotou, C., Meegan, C.A., Fishman, G.J. et al. 1993, \apjl, 413, 101
\bibitem[{Mangano et al. }{2008}]{man08} Mangano V., Sbarufatti, B., Ukwatta, T.N., et al. 2008,GCN 7208
\bibitem[{M\'{e}sz\'{a}ros \& Rees }{1993}]{mer93} M\'{e}sz\'{a}ros P \& Rees M. 1993, \apjl 405, 278
\bibitem[{Metzger et al., }{2008}]{met08} Metzger, B.D., Quataert, E. Thompson, T. A. 2008, \mnras, 385, 1455
\bibitem[{Nakar }{2007}]{nak07} Nakar E. 2007, PhR, 442, 166
\bibitem[{Nakar et al. }{2009}]{nak09} Nakar E., Ando, S., Sari, R. 2009, arXiv:0903.2557
\bibitem[{Norris \& Bonnell }{2006}]{nob06} Norris, J.P., \& Bonnel, J.T. 2006, \apj, 643, 266
\bibitem[{Oates et al. }{2009}]{oat09} Oates, S.R., Page, M. J., Schady, P. et al. 2009 \mnras 395, 490
\bibitem[{Ohno, \& Pelassa }{ 2009}]{ohp09} Ohno, M. \& Pelassa, V. 2009, GCN 9334
\bibitem[{Panaitescu et al. }{2006}]{pan06} Panaitescu A., M\'{e}sz\'{a}ros, P., Burrows, D., et al. 2006, \mnras, 369, 2059
\bibitem[{Panaitescu \& Verstrand }{2008}]{pav08} Panaitescu A. \& Vestrand T.  2008, \mnras, 387, 479.
\bibitem[{Perna et al. }{2006}]{per06} Perna, R. Armitage, P. J., Zhang, B. 2006, \apjl, 636, 29
\bibitem[{Poole et al. }{2008}]{poo08} Poole, T.S., Breeveld, A.A., Page, M.J., et al. 2008, \mnras, 383, 627 
\bibitem[{Rau et al. }{2009}]{rau09} Rau A., McBreen S., Kruehler T. et al. 2009, GCN 9353
\bibitem[{IS, Rees \& M\'{e}sz\'{a}ros }{1994}]{rem94} Rees, M. J. \& Meszaros, P. 1994, \apjl, 430, 93
\bibitem[{UVOT; 160-800~nm; Roming et al. }{2005}]{rom05} Roming, P.W.A., Kennedy, T.E., Mason, K.O., et al. 2005, SSRv, 120, 95
\bibitem[{Rosswog }{2007}]{ros07}Rosswog, S. 2007, RMxAC, 27, 57R
\bibitem[{Sari }{1997}]{sar97} Sari, R. 1997, \apjl, 489, 37
\bibitem[{Sari \&  M\'{e}sz\'{a}ros }{2000}]{sam00} Sari, R., \& M\'{e}sz\'{a}ros P. 2000, \apjl, 535, 33
\bibitem[{Sari et al. }{1998}]{sar98} Sari, R., Piran, T., \& Narayan R. 1998, \apjl, 497, 17
\bibitem[{Sari et al. }{1999}]{sar99} Sari, R., Piran., T., and Helpern, J. 1999, \apjl, 519, 17
\bibitem[{Stratta et al. }{2007}]{stra07} Stratta, G., D'Avanzo, P., Piranomonte, S. et al. 2007, A\&A, 474, 827
\bibitem[{Troja et al. }{2008}]{tro08}
Troja, E., King, A.R., O'Brien, P.T. et al. 2008, \mnras, 385, 10.
\bibitem[{Ukwatta et al. }{2009}]{ukw09} Ukwatta., T.N., Barthelmy, S.D., Baumgartner W.H. et al. 2009, GCN 9337
\bibitem[{Yamazaki et al. }{2002}]{yam03} Yamazaki, R., Ioka, K., \& Nakamura, T. 2002, \apjl, 571, 31
\bibitem[{Zhang et al. }{2009}]{zha09} Zhang, B., Zhang, B.-B.,
Virgili, F. J. et al. 2009, \apj~in press, arXiv:0902.2419.
\end{thebibliography}
\end{document}